FRONT MATTER

## Title
- Controlling Electromagnetic Surface Waves with Conformal Transformation Optics

## Authors

Xiaoyu Zhao,[1]† Hong Deng,[1]† Xiaoke Gao,[1]† Xikui Ma,[1] Tianyu Dong[1]*

[1]School of Electrical Engineering, Xi'an Jiaotong University, Xi'an 710049, China.

†These authors contributed equally to this work.
*To whom correspondence should be addressed; E-mail: tydong@mail.xjtu.edu.cn.

## Abstract

The application of transformation optics to the development of intriguing electromagnetic devices can produce weakly anisotropic or isotropic media with the assistance of quasi-conformal and/or conformal mapping, as opposed to the strongly anisotropic media produced by general mappings; however, it is typically limited to two-dimensional applications. By addressing the conformal mapping between two manifolds embedded in three-dimensional space, we demonstrate that electromagnetic surface waves can be controlled without introducing singularity and anisotropy into the device parameters. Using fruitful surface conformal parameterization methods, a near-perfect conformal mapping between smooth manifolds with arbitrary boundaries can be obtained. Illustrations of cloaking and illusions, including surface Luneburg and Eaton lenses and black holes for surface waves, are provided. Our work brings the manipulation of surface waves at microwave and optical wavelengths one step closer.

## Teaser
Waves can be controlled at will on arbitrary open surfaces without holes, showing fascinating applications such as invisible bumps for surface waves, reproducing scatterings of one bump on other smooth surfaces, and controlling light beams on surfaces to focus, to bend and/or to be absorbed akin to black holes without visible scatterings.

# MAIN TEXT

## Introduction

Since its inception in the design of electromagnetic cloaks (*1*, *2*), transformation optics (TO) has proven to be a powerful tool for understanding and customizing the physics in acoustics (*3*), optics (*4*), mechanics (*5*), thermodynamics (*6*, *7*), *etc*. Following the groundbreaking work of cloaking, a number of other electromagnetic devices have been reported within the theoretical framework of TO, such as electromagnetic concentrators (*8*, *9*), field rotators (*10*), optical lenses (*11*, *12*) and optical illusion devices (*13*, *14*). In practice, however, traditional TO often yields significant anisotropy in a designed medium (*15*). Thus, metamaterials are often used to infer spatial changes from coordinate transformation geometry, which is based on the mathematical equivalence between geometry and material (*16*).

To reduce the anisotropy of the functional medium induced by TO, various approaches have been developed. By constructing mapping in non-Euclidean space, for instance, it is possible to remove singular points formed by traditional TO (*17*), hence minimizing anisotropy in part. But for wavelengths comparable to the size of the transform region, non-Euclidean TO may perform even worse (*18*); thus, several research projects focus on conformal or quasi-conformal mappings to achieve isotropy (*19*). In $\mathbb{R}^2$, the concept of a carpet cloak that resembles a flat ground plane is successfully realized with an isotropic medium produced by minimizing the Modified-Liao functional under sliding boundary conditions (*20*), or equivalently by constructing the quasi-conformal mapping via solving inverse Laplace's equations (*21*). Although the concept of carpet cloak has been extended to $\mathbb{R}^3$ by the extrusion or revolution of a two-dimensional refractive index profile to control the reflection of free-space waves, it is only applicable to surfaces with translational or rotational symmetry (*22*).

Previous research has focused largely on controlling propagating waves by TO, whereas less attention has been attached to the manipulation of surface waves (*12*, *23*, *24*). Perfect surface wave cloaks have been proposed by equating the optical path length of a ray traversing a flat plane with a homogeneous refractive index to the optical path on a curved surface with an angle-dependent refractive index for two orthogonal paths (*25*, *26*), which have been experimentally validated (*27*). Although an electrically large object may be hidden by such a cloak with an inhomogeneous isotropic medium, this approach is limited to rotationally symmetric surfaces. By linking the governing eikonal equations on a virtual flat plane and on a curved surface by transformation optics, the projection mapping yields surface wave cloaks for non-rotationally symmetric geometries but with high anisotropy (*14*, *28*). Considerable effort has been devoted to reducing such anisotropy by employing efficient numerical conformal algorithms such as boundary first flattening (*29*), yet only non-rotationally symmetric surfaces with circular boundary are investigated (*30*).

In this work, we show how to manipulate surface waves on smooth manifolds embedded in $\mathbb{R}^3$ within the framework of conformal TO, requiring an effective isotropic material under the regime of geometrical optics. Fig. 1 illustrates a conformal surface mapping between two smooth manifolds in $\mathbb{R}^2$ and $\mathbb{R}^3$, *i.e.*, $f: \mathcal{M}' \to \mathcal{M}$. The curved manifold $\mathcal{M}$ shown in Fig. 1A has been $uv$-parameterized and the mesh grid can be regarded as the mapping result of the Cartesian coordinate system $\{x', y'\}$ in Fig. 1B. When the mapping is conformal or quasi-conformal, the face element $dS$ remains right-angled, indicating that elements are just scaled up with little distortion. From the local coordinate systems on $dS$ and $dS'$ (Fig. S5), one can derive the Jacobian matrix $\mathbf{J}$ of mapping $f$ with two singular values $\sigma_{J1} = \sigma_{J2} = \sigma_J$ that state equal scaling in two orthogonal directions (*31*). Consequently, an isotropic cloaking medium distribution $n = 1/\sqrt{\det(\mathbf{J})} = 1/\sigma_J$ may be obtained based on the conformal TO (*19*), representing the ratio of line element $dl'$ in virtual space to the scaled element $dl$ in physical space for compensating optical path length (*2*). As a result, light propagating on curved $\mathcal{M}$ behaves as propagating on flat $\mathcal{M}'$. In practice, it is more convenient to describe mesh vertices in $\mathbb{R}^3$ in a Cartesian coordinate system $\{x, y, z\}$ and the Jacobian derived from the local coordinate system forms an asymmetric rank-two matrix $\mathbf{J}_{3\times 2}$. In addition, the possible quasi-conformal mappings can be measured by the conformality, *i.e.*, the ratio $Q = \max(\sigma_{J1}/\sigma_{J2}, \sigma_{J2}/\sigma_{J1})$. A unity ratio $Q$ allows an effective cloaking medium expressed as $n_{\text{cloak}} = 1/\sqrt{\sigma_{J1}\sigma_{J2}}$ for every face element (*20*).

## Results

Having obtained a conformal mapping between the manifolds $\mathcal{M} \in \mathbb{R}^3$ and $\mathcal{M}' \in \mathbb{R}^2$, we first design an isotropic surface wave cloak under the perspective of conformal TO and compare its performance with the traditional surface wave cloak with anisotropic medium (*14*). Simulations were conducted on a double-camelback bump with an elliptical base profile embedded in $\mathbb{R}^3$, as shown in Fig. 2. In comparison with the scattering when the surface has no index profile (Fig. S1A), one can observe that the surface wave cloaking is successfully achieved by two distinct approaches: one induced by the projection mapping proposed in (*14*) (Fig. 2A) and the other originated from the proposed quasi-conformal mapping (Fig. 2B). The corresponding material characteristics for the two types of cloaks are displayed in Fig. 2C, indicating that the former is strongly anisotropic while the latter is almost isotropic. In addition, the isotropic refractive index $n_{c,\text{double}}$ (the subscript "c" denotes the cloak, and "double" denotes the double-camelback bump) ranges from 0.83 to 1, which decreases as the bump height rises because a longer geometrical distance need to be compensated by a smaller refractive index in order to attain equal optical path length.

The proposed scheme based on conformal TO has achieved near-perfect surface wave cloaking while eliminating the anisotropy in the transformation medium that the traditional scheme presents. The distribution of $n_{c,\text{double}}$ in Fig. 2C outlines an asymmetric geometric profile, manifesting that the effectiveness of this scheme is independent from any symmetry. Such an achievement demands mappings with high conformality rather than those bringing large distortion such as the projection mapping (*14*). The numerical method we adopt here (*29*) can obtain a quasi-conformal mapping with $Q < 1.03$, as shown in Fig. S1B, which is sufficient for designing an effective isotropic cloaking medium distribution.

As the antithesis of cloaking, optical illusion devices can reproduce the scattering characteristics of a specific object on other objects through a transformation medium (*13, 14*). Fig. 3A depicts the surface electromagnetic wave scattered by a single-camelback bump $\mathcal{M}$ filled with homogeneous material. Traditionally, if one wants to reproduce its scattering on a plane region $\mathcal{M}'$, the quasi-conformal mapping for designing the illusion device is $f': \mathcal{M} \to \mathcal{M}'$ with a Jacobian matrix $\Lambda_{2\times 3}$. Fig. 3B shows the accurately recurring scattering characteristics on plane region $\mathcal{M}'$ filled with $n_{i,\text{plane}} = 1/\sqrt{\sigma_{\Lambda 1}\sigma_{\Lambda 2}}$ (the subscript "i" denotes the illusion, and "plane" denotes the plane region), where $\sigma_{\Lambda 1}$ and $\sigma_{\Lambda 2}$ are singular values of $\Lambda_{2\times 3}$. Furthermore, Fig. 3C illustrates that the double-camelback bump filled with a carefully designed isotropic medium distribution can reproduce the same scattering pattern as shown in Fig. 3A. Such an illusion is realized by cascading two conformal mappings described in Fig. S3, *i.e.*, $f_1$ from $\mathbb{R}^3$ (virtual space) to $\mathbb{R}^2$ (intermediate space), and $f_2$ from $\mathbb{R}^2$ to the $\mathbb{R}^3$ (physical space). Thus, the illusion medium for the double-camelback bump reads $n_{i,\text{double}} = n_{i,\text{plane}} \cdot n_{c,\text{double}}$. Fig. 3D displays the profiles of $n_{i,\text{plane}}$ (for Fig. 3B) and $n_{i,\text{double}}$ (for Fig. 3C), respectively, which range from 1 to 1.25 ($n_{i,\text{plane}}$) and from 0.85 to 1.21 ($n_{i,\text{double}}$).

The scattering pattern of the single-camelback bump (Fig. 3A) has been successfully reproduced on the plane region (Fig. 3B) and on the double-camelback bump (Fig. 3C), which demonstrates that the proposed scheme is a general solution to illusion design on smooth two-dimensional manifolds. The cascading method to construct mappings between manifolds embedded in $\mathbb{R}^3$ can even tackle surfaces with different base profiles, since a conformal mapping between simply-connected regions in $\mathbb{R}^2$ exists according to the Riemann mapping theorem (*32*). Moreover, the quasi-conformal ratios $Q$ of the two mappings for the double-camelback and single-camelback bump are smaller than 1.03 (Fig. S1B) and 1.012 (Fig. S2C), respectively, implicating that the cascaded mapping meets the requirement for high conformality. The range of $n_{i,\text{single}}$ (1 to 1.25) is the inverse of that of the cloaking refractive index $n_{c,\text{single}}$ (0.8 to 1) shown in Fig. S2B, because the illusion can be regarded as the inverse design of the cloaking such that the Jacobian matrices of their corresponding mappings are the Moore–Penrose pseudo-inverse of each other (*31*).

Now that the wave behavior on the curved manifold can be manipulated flexibly, it is natural to consider designing various complicated devices on it, such as surface wave Luneburg lens, Eaton lens and black hole for surface waves (*12, 23, 33, 34*). Traditional designs are usually based on spherical or circular profiles with a constant radius. While for an elliptical profile without a constant radius, we adopt the distance from the point on the ellipse to the center, also the coordinate origin, as the generalized radius, *i.e.*, $R(\theta) = \sqrt{(a\cos\theta)^2 + (b\sin\theta)^2}$ (*35–37*). Thus, the refractive index of the considered Luneburg lens can be expressed as

$$n_\text{L}(r,\theta) = \sqrt{2 - (r/R(\theta))^2}, \tag{1}$$

where $r = \sqrt{x^2 + y^2}$ and $\theta = \arctan(y/x)$. Similar to the traditional circular Luneburg lens, such a distribution retains $n_\text{L} = 1$ on the boundary and $n_\text{L} = \sqrt{2}$ at the center $r = 0$ (*38*). Next, the medium distribution for a Luneburg lens on the double-camelback bump can be expressed as $n_\text{Luneburg} = n_{c,\text{double}} \cdot n_\text{L}$. As illustrated in Fig. 4A, two Gaussian beams with a free-space wavelength $\lambda_\text{G} = 50$ mm are incident along the $x$-direction at the position $\pm 0.8b$ on the $y$-direction and reflected by the Luneburg lens to interfere at the focus point. The focal distance reads $20\lambda_\text{G}$ that is identical to the unit circular Luneburg lens. For the Eaton lens, the refractive index $n_\text{E}$ reads as

$$n_\text{E}(r,\theta) = \sqrt{2R(\theta)/r - 1}, \tag{2}$$

which can approach infinity when $r = 0$, leaving a singular point to be cared for. Fig. 4B describes that a Gaussian beam going along the $x$-direction bends to the inverse $x$-direction after passing through the Eaton lens on the double-camelback bump. The proposed surface wave Luneburg and Eaton lenses may be deployed in optical imaging, signal acquisition and novel designs for surface wave microwave antennas. Another functional device that can rotate beam propagation is the peripheral of the two-layer optical black hole, where light is compelled to travel in a spiral path into the absorbing medium at the core. The piece-wise refractive index distribution function $n_\text{B}$ can be expressed as

$$n_\text{B}(r,\theta) = \begin{cases} 1, & r > R(\theta) \\ R(\theta)/r, & r_c \cdot R(\theta) < r < R(\theta), \\ 1/r_c + i\gamma, & r < r_c \cdot R(\theta) \end{cases} \tag{3}$$

where $r_c = 0.4$ is the scaling factor of the internal ellipse core compared with the base profile and $\gamma = 0.1$ is the loss factor. The refractive index distribution $n_\text{Blackhole} = n_{c,\text{double}} \cdot n_\text{B}$ on the double-camelback bump is depicted in Fig. 4D. The real part of material parameters is matched on the inner boundary, and the imaginary part for absorbing energy ranging from 0.083 to 0.097 only exists in the core. The same Gaussian beam that was used for the Eaton lens is employed, and the result in Fig. 4C shows that the beam bends around 90° before it reaches the inner boundary and is absorbed by the lossy core without reflection, showing potential application in interference reduction and energy harvesting for electronic devices. Note that, the overall sizes of the simulation models are larger than ten times the operating wavelength, demonstrating that the proposed scheme is capable of managing surface wave behaviors on electrically large objects. Moreover, the excellent performance of these functional surface wave devices demonstrates that, based on the proposed scheme, a variety of novel devices may be realized on smooth curved manifolds, which may facilitate the development of miniaturized and integrated photonic devices.

**Discussion**

Our theory and method are based on geometrical optics. It requires small curvature and little variation in wavelength (see (7) and (8) in *Materials and Methods*), which can be expressed as

$$w = |\nabla \lambda| = |\nabla(\lambda_0/n)| = \lambda_0 |\nabla n|/n^2 \ll 1, \tag{4}$$

$$\rho = |R_{ij}|\lambda^2 = |K g_{ij}|(\lambda_0/n)^2 = \det(g_{ij}) K^2 \lambda_0^2/n^2 = K^2 \lambda_0^2/n^6 \ll 1, \tag{5}$$

where $R_{ij}$ is the Ricci curvature tensor, $K$ is the Gaussian curvature, and $g_{ij}$ is the metric tensor. Both the wavelength index $w$ and the curvature index $\rho$ are inversely proportional to powers of the refractive index $n$. In order to prevent $w$ and $\rho$ from increasing drastically, a height lower than half of the base radius is favorable, and thereby the optical path length can be compensated with a near-unity refractive index. On this basis, requirements (4) and (5) demand shorter wavelength $\lambda_0$ and smoother geometric structure to ease the changing rate $|\nabla n|$ and the Gaussian curvature $K$. As a negative example, a hemisphere surface wave cloak is reviewed and results are displayed in Fig. S4, whose refractive index $n_{c,\text{sphere}}$ is between 0.5 and 1 and the maximum of quasi-conformal ratio $Q$ is smaller than 1.012. The visible scattering appearing in Fig. S4C implies the failure of geometrical optics because of the high curvature index $\rho > 20$ residing in the right-angle connection between the hemisphere and the plane, as is depicted in Fig. S4F, and the average curvature index $\bar{\rho} = 1.57$ is also larger than 1. The non-smooth connection causes the phase distortion in the backward scattering, and the maximum of the forward scattering $|E_z - E_{bz}|_{\max} = 0.75$ V/m implies a phase difference $\arcsin(0.75) = 48.6°$ resulted from the reconstruction of wave fronts. In comparison, Fig. S1C and Fig. S2D display the average curvature index $\bar{\rho} = 0.54$ for double-camelback bump and $\bar{\rho} = 0.39$ for single-camelback bump, respectively, both satisfying the requirement (5) and leaving near-zero $\rho$ on smooth boundaries. One may notice that the wavelength index $w$ for the cloaks shown in Fig. S1D, Fig. S2E and Fig. S4E is smaller than unity everywhere because it is related to lower powers of $\lambda_0$ and $n$; thus, it is much easier to meet the requirement of (4) compared to (5). These selected curvature and wavelength characteristics that validate the approximation of geometrical optics are indispensable for the excellent performance of electromagnetic devices.

The isotropic case that determines the expression of requirements (4) and (5) is based on the conformal or quasi-conformal mappings between two-dimensional manifolds. Benefiting from the rapid development in conformal parameterization, a series of mapping methods can be employed to design surface wave carpet cloak (*29*, *39*, *40*). The boundary first flattening (BFF) method (*29*) adopted in our study can establish near-perfect conformal mappings not only between smooth manifolds but also surfaces with cuspidal points, such as sharp corners and cone singularities, offering exhilarating promise for wave manipulation on more complicated surfaces. In addition, there are algorithms aimed at constructing quasi-conformal mappings between high-genus manifolds (*41*, *42*), which can be used to deal with phase regulation on surfaces with holes. One noteworthy idea is to map a high-genus surface to a zero-genus plane region by transforming holes to slits (*43*, *44*) that implies the possibility for the scheme conducted in simply-connected regions to manipulate wave behaviors on multiply-connected surfaces. By reasonably utilizing advanced algorithms for a variety of particular cases, our method has the potential to be a universal scheme for controlling surface electromagnetic waves on an arbitrary two-dimensional manifold.

In summary, we have proposed a general method to manipulate electromagnetic waves on smooth two-dimensional manifolds without rotational symmetry by means of a certain isotropic refractive index distribution derived from the quasi-conformal mapping. The relationship between medium and mappings is induced from the wave equation on the manifold under the geometrical optics approximation. Numerical quasi-conformal algorithms are introduced to construct mappings between manifolds, and consequent functional mediums are validated by cloaking surfaces and generating illusions on plane regions. By cascading mappings between $\mathbb{R}^2$ and $\mathbb{R}^3$ to obtain a mapping between $\mathbb{R}^3$, we succeed in reproducing the scattering of a surface on another surface. In addition, functional devices such as surface Luneburg lenses, surface Eaton lenses, and black holes for surface waves are designed based on carpet cloaks. Finally, the indices required by geometrical optics are reviewed to demonstrate the validity of the approximation on simulation models. Our method paves the way for the regulation of surface electromagnetic waves on any two-dimensional manifold, and can be utilized to control surface waves in other fields, such as acoustics, mechanics, and thermodynamics.

## Materials and Methods

### Conformal transformation optics for surface waves

**Wave equation on curved manifold.** The concept of transformation medium stems from the equivalence between geometry and media. Within the Einstein summation convention, the Maxwell's wave equation for the electric field $\nabla_\mathcal{M} \times \nabla_\mathcal{M} \times \mathbf{E} - \mu_0 \varepsilon_0 \partial_t^2 \mathbf{E} = 0$ in free space can be expressed as (*16*)

$$\nabla^j \nabla_j E_i - R_{ij} E^j - c_0^{-2} \partial_t^2 E_i = 0, \tag{6}$$

where $c_0 = 1/\sqrt{\mu_0 \varepsilon_0}$ is the light velocity in free space; $R_{ij}$ is the Ricci tensor of the considered geometry $\mathcal{M}$. Supposing that the electromagnetic waves are confined nearby a curved surface $\mathcal{M}$ embedded in $\mathbb{R}^3$ as surface waves, its local plane wave solution reads as $E_i = \mathcal{E}_i e^{i\varphi}$ with constant complex amplitudes $\mathcal{E}_i$, where the phase reads as $\varphi = \mathbf{k} \cdot \mathbf{r} - \omega t$ with the wave vector $\mathbf{k} = \nabla_\mathcal{M} \varphi$ and angular frequency $\omega = -\partial_t \varphi$. For surface waves, the wave vector k lies in the tangent space of the curved surface $\mathcal{M}$, i.e., $\mathbf{k} \in \mathcal{T}(\mathcal{M})$. Thus, (6) can be simplified and approximated in the regime of geometrical optics where the wavelength $\lambda = 2\pi/k$ varies slowly with distance, i.e.,

$$|\nabla_\mathcal{M} \lambda| \ll 1. \tag{7}$$

In addition, the effective curvature of the curved surface should be small enough compared to the wavelength so that the assumption of locally plane waves is valid, i.e.,

$$|R_{ij}|\lambda^2 \ll 1. \tag{8}$$

As a result, inserting $E_i = \mathcal{E}_i e^{i\varphi}$ into (6) and considering that the (spatial and temporal) derivatives of $\mathcal{E}_i$ vanish, one can obtain the dispersion relation for the surface wave propagating on $\mathcal{M}$, which reads as

$$k^2 = k^j k_j = g^{ij} k_i k_j = \omega^2/c_0^2. \tag{9}$$

Here, $g_{ij}$ is the induced metric tensor for the curved surface $\mathcal{M}$, which can be determined from the transformation Jacobian matrix from the manifold $\mathcal{M}'$ in $\mathbb{R}^2$ to $\mathcal{M}$ (*31*),.

**Wave equation on a flat plane.** Alternatively, if $\mathcal{M}$ is flat (i.e., $R_{ij} = 0$) and filled with anisotropic medium denoted by relative permeability tensor $\mu^{ij}$, (6) becomes

$$\nabla \times \nabla \times \mathbf{E} - \mu_0 \varepsilon_0 \boldsymbol{\mu} \cdot \partial_t^2 \mathbf{E} = 0. \tag{10}$$

Suppose that the electromagnetic waves are confined nearby $\mathcal{M}$ and the electric field $\mathbf{E}$ is perpendicularly polarized. In a Cartesian coordinate system, if $\mathcal{M}$ can be placed into $xy$ plane, we focus on the case that the electric field vector $\mathbf{E}$ lies in the normal space of the flat plane $\mathcal{M}$, i.e., $\mathbf{E} \in \mathcal{N}(\mathcal{M})$, and the global wave solution may read as $E_z = \mathcal{E}_z e^{i\varphi}$. Thus, the phase $\varphi$ is independent of $z$ and the wave vector just lies on the plane as $\mathbf{k} = (k_x, k_y, 0)$, because a flat plane is coincident with its tangent space. Since the flat manifold $\mathcal{M}$ has a zero-curvature tensor, the condition (7) holds naturally. Once the other condition (8) that wavelength varies slowly is satisfied, one may disregard the derivatives of complex amplitude after inserting $E_z = \mathcal{E}_z e^{i\varphi}$ into (10) and obtain the dispersion relation for the surface wave propagating on $\mathcal{M}$, which reads as $(\mu_{xx} k_x^2 + 2\mu_{xy} k_x k_y + \mu_{yy} k_y^2)/\det(\boldsymbol{\mu}) = \omega^2/c_0^2$. By excluding consideration of the particular polarization, the dispersion equation can be recast within the Einstein summation convention as

$$\frac{1}{\det(\boldsymbol{\mu})} \mu^{ij} k_i k_j = \frac{\omega^2}{c_0^2}. \tag{11}$$

**Transformation medium and geometry.** For electromagnetic waves that behave identically on two manifolds, one can obtain the equivalence between geometry and material properties by comparing (9) and (11), which yields

$$\frac{\mu^{ij}}{\det(\boldsymbol{\mu})} = g^{ij}. \tag{12}$$

The relative permeability tensor $\mu^{ij}$ actually creates an illusion on the flat plane because a spatial point filled with medium $\boldsymbol{\mu}$ is equivalent to be with a metric $\boldsymbol{g} = \det(\boldsymbol{\mu})\boldsymbol{\mu}^{-1}$. If the local Cartesian coordinate system at this point is aligned along the orthogonal eigenvectors of $\boldsymbol{\mu}$, the real and symmetric permeability tensor will reduce to diag$(\mu_x, \mu_y, \mu_z)$ so that the square of the line element on $x$ direction is $ds^2 = g_{xx} dx^2 = \mu_y \mu_z dx^2$, which is also the square of optical path length in curved free space. In comparison to $ds^2 = n_x^2 dx^2$ on the flat manifold, one can derive $n_x^2 = \mu_y \mu_z$ and similar results on $y$ and $z$ directions. Consequently, the relationship between the relative permeability tensor $\boldsymbol{\mu}$ and the refractive index tensor $\boldsymbol{n}$ may be expressed as $\boldsymbol{n}^2 = \det(\boldsymbol{\mu})\boldsymbol{\mu}^{-1}$ and one may further obtain

$$\boldsymbol{n}_{\text{illustion}}^2 = \boldsymbol{g}. \tag{13}$$

by referring to (12).

**Surface transformation and TO medium.** The metric tensor in equation (13) is induced from the mapping $f: \mathcal{M}' \to \mathcal{M}$ and can be constructed by the Jacobian matrix $\mathbf{J}_{3\times 2}$ as $\boldsymbol{g} = \mathbf{J}^T \mathbf{J}$ (*31*). Nevertheless, we prefer to associate $\boldsymbol{n}_{\text{illusion}}$ with the Jacobian matrix $\boldsymbol{\Lambda}_{2\times 3}$ that represents the transformation from $\mathbb{R}^3$ (virtual space) to $\mathbb{R}^2$ (physical space). Actually, the asymmetric Jacobian matrices $\mathbf{J}_{3\times 2}$ and $\boldsymbol{\Lambda}_{2\times 3}$ can be denoted as the Moore–Penrose pseudo-inverse of each other (*31*), *i.e.*, $\mathbf{J} = \boldsymbol{\Lambda}^\dagger$, where the superscript '†' denotes pseudo-inverse. Thus, one can rewrite the equivalence (13) as

$$\boldsymbol{n}_{\text{illustion}}^2 = \boldsymbol{g} = \mathbf{J}^T \mathbf{J} = (\boldsymbol{\Lambda}\boldsymbol{\Lambda}^T)^{-1}. \tag{14}$$

Similar relationship can be obtained for cloaking medium $\boldsymbol{n}_{\text{cloak}}$ and corresponding Jacobian matrix $\mathbf{J}_{3\times 2}$ from $\mathbb{R}^2$ (virtual space) to $\mathbb{R}^3$ (physical space) as

$$\boldsymbol{n}_{\text{cloak}}^2 = (\mathbf{J}^T \mathbf{J})^{-1}. \tag{15}$$

For the mapping between $\mathbb{R}^3$ (Fig. S3), which is formed by cascading two transformations between $\mathbb{R}^3$ and $\mathbb{R}^2$, the consequent medium for the illusion can be recast as the combination of the cloaking and illusion refractive index tensors, *i.e.*,

$$\boldsymbol{n}_{\text{illustion}}^2 = (\boldsymbol{\Lambda}_1 \boldsymbol{\Lambda}_1^T)^{-1} \cdot (\mathbf{J}_2^T \mathbf{J}_2)^{-1}. \tag{16}$$

where $\boldsymbol{\Lambda}_1$ and $\mathbf{J}_2$ are Jacobian matrices for mappings $f_1$ and $f_2$, as illustrated in Fig. S3, respectively. In particular, when the mappings are conformal, the refractive index becomes isotropic, and the corresponding Jacobian matrix has two identical singular values. By taking the determinants of (14) and (15), the refractive indices can be denoted by singular values of Jacobian matrices as $n_{\text{cloak}} = 1/\sigma_J$ and $n_{\text{illusion}} = 1/\sigma_\Lambda$.

### Discrete conformal mapping and transformation medium

**Review on discrete conformal mapping.** It has been demonstrated that an isotropic refractive index distribution can be achieved by solving equations for equal optical path length only on rotationally-symmetric surfaces (*25*). As to the non-rotationally symmetric cloak, high anisotropy is introduced by the projection mapping that distorts the coordinate grid (*14*). However, numerical algorithms for surface parameterization provide possible conformal mappings for arbitrary surfaces. For example, the angle-based flattening (ABF) method (*45*, *46*) has been proposed to construct conformal parameterization by minimizing a punishing functional to decrease angular distortion while its nonlinearity reduces computational efficiency. Also, the so-called least-squares method (LSCM) (*47*) and spectral method (SCP) (*48*) have been introduced to attain higher efficiency, benefiting from their linearity. Their disadvantages are free target boundaries and non-bijectivity, whereas we expect a one-to-one mapping that includes every point on physical and virtual space with controlled boundaries. Further research, like disk conformal mapping (DCM) (*40*), has been reported as a linear and bijective conformal mapping method but with a fixed disk boundary. Not until boundary first flattening (BFF) (*29*) enabled editing boundary as demand were the drawbacks totally eliminated. To deal with a certain electromagnetic circumstance, one could choose an appropriate algorithm among the preceding techniques (*49*, *50*).

**Triangulation and Jacobian matrices.** Supposing that the conformal mapping reads $f_1: \mathcal{M}_2 \to \mathcal{M}_1$ (or $f_2: \mathcal{M}_1 \to \mathcal{M}_2$) between manifolds $\mathcal{M}_1 \subset \mathbb{R}^3$ and $\mathcal{M}_2 \subset \mathbb{R}^2$, as shown in Fig. S5A, one can find that a simplex $\mathcal{S}_1$ on meshed $\mathcal{M}_1$ and its counterpart on meshed $\mathcal{M}_2$ are a pair of similar triangles, which allows $\mathcal{S}_1$ and $\mathcal{S}_2$ to share a same barycentric coordinate system. This local coordinate system, as shown in Fig. S5B, can represent any point inside the simplex as the linear combination of three vertices and helps quickly induce the Jacobian matrix of numerical mappings based on triangular mesh parameterization. For example, the location of the point $\mathbf{q}(x', y')$ on $\mathcal{S}_2$ can be expressed as $x' = \sum_{i=1}^{3} \lambda_i x_i'$ and $y' = \sum_{i=1}^{3} \lambda_i y_i'$ with $\lambda_1 + \lambda_2 + \lambda_3 = 1$, *i.e.*, a linear combination of vertices $\mathbf{q}_1(x_1', y_1')$, $\mathbf{q}_2(x_2', y_2')$ and $\mathbf{q}_3(x_3', y_3')$. For the triangulation mesh, we can obtain the barycentric coordinates, which read as

$$\lambda_1 = [(y_2' - y_3')(x' - x_3') + (x_3' - x_2')(y' - y_3')]/\det(\mathbf{M}), \tag{17}$$

$$\lambda_2 = [(y_3' - y_1')(x' - x_3') + (x_1' - x_3')(y' - y_3')]/\det(\mathbf{M}), \tag{18}$$

$$\lambda_3 = [(y_1' - y_2')(x' - x_2') + (x_2' - x_1')(y' - y_2')]/\det(\mathbf{M}), \tag{19}$$

where $\det(\mathbf{M}) = |(\mathbf{q}_1 - \mathbf{q}_3) \times (\mathbf{q}_2 - \mathbf{q}_3)|$, with $\mathbf{q}_i(x_i', y_i')$ being the $i$-th vertices ($i = 1, 2, 3$). Here, (17), (18) and (19) show that the barycentric coordinate system $(\lambda_1, \lambda_2, \lambda_3)$ can be expressed by the Cartesian coordinate system $(x', y')$. Regarding the point $\mathbf{p}(x, y, z)$ on $\mathcal{S}_1 \subset \mathbb{R}^3$, mapped from the point $\mathbf{q}$ in $\mathbb{R}^2$, we have $x = \sum_{i=1}^{3} \lambda_i x_i$, $y = \sum_{i=1}^{3} \lambda_i y_i$ and $z = \sum_{i=1}^{3} \lambda_i z_i$ as the linear combination of $\mathbf{p}_1(x_1, y_1, z_1)$, $\mathbf{p}_2(x_2, y_2, z_2)$ and $\mathbf{p}_3(x_3, y_3, z_3)$, since $\mathcal{S}_1$ and $\mathcal{S}_2$ share the same barycentric coordinates $\lambda_i$. As a result, the Jacobian matrix $\mathbf{J}_{3\times 2}$ of the mapping from $\mathcal{S}_2 \subset \mathbb{R}^2$ to $\mathcal{S}_1 \subset \mathbb{R}^3$ can be derived according to the derivatives of $(\lambda_1, \lambda_2, \lambda_3)$ with respect to $(x', y')$, which reads as

$$\mathbf{J}_{3\times 2} = \begin{pmatrix} \partial_{x'} x & \partial_{y'} x \\ \partial_{x'} y & \partial_{y'} y \\ \partial_{x'} z & \partial_{y'} z \end{pmatrix} = \frac{1}{\det(\mathbf{M})} \begin{pmatrix} x_1 & x_2 & x_3 \\ y_1 & y_2 & y_3 \\ z_1 & z_2 & z_3 \end{pmatrix} \begin{pmatrix} y_2' - y_3' & x_3' - x_2' \\ y_3' - y_1' & x_1' - x_3' \\ y_1' - y_2' & x_2' - x_1' \end{pmatrix}. \tag{20}$$

In a similar manner, one can derive the Jacobian matrix $\mathbf{\Lambda}_{2\times 3}$ of the numerical mapping from $\mathcal{S}_1$ to $\mathcal{S}_2$; alternatively, one may calculate the Moore–Penrose pseudoinverse of $\mathbf{J}_{3\times 2}$ as $\mathbf{\Lambda}_{2\times 3}$ (*31*). By calculating the Jacobian matrices $\mathbf{J}_{3\times 2}$ or $\mathbf{\Lambda}_{2\times 3}$ on each simplex, the information of mapping $f_1$ or $f_2$ can be fully described.

**Simulation methods**

**FEM simulation.** The wave behavior of electromagnetic devices is simulated using the finite element method. The geometric model is an optical thin-film waveguide whose thickness is less than one fifth of the wavelength. On the outer surfaces of the waveguide, the perfect electric conductor (PEC) boundary condition is applied to emulate the propagation of the surface wave on a two-dimensional manifold. Thus, the propagation of the plane wave or Gaussian beam is restricted within the optical thin film. To mimic an open and non-reflecting infinite domain, perfectly matched layers (PMLs) are applied on the boundary of the propagating plane. The designed medium is configured to the waveguide as a fitting function interpolated from the discrete data set calculated on extra dense meshes.

**Acknowledgments**

**Funding:** This work was supported by the National Natural Science Foundation of China (NSFC) under grant no. 51977165.

**Author contributions:**

Conceptualization: X.Z., X.M., T.D.

Methodology: X.Z., H.D., X.G., T.D.
Investigation: X.Z., H.D., X.G.
Visualization: X.Z., H.D., X.G.
Supervision: X.M., T.D.
Writing—original draft: All authors.
Writing—review & editing: All authors.

**Competing interests:** The authors declare that they have no competing interests.

**Data and materials availability:** All data needed to evaluate the conclusions in the paper are present in the paper and/or the Supplementary Materials. Raw data and corresponding simulation data are available upon request.

**Figures and Tables**

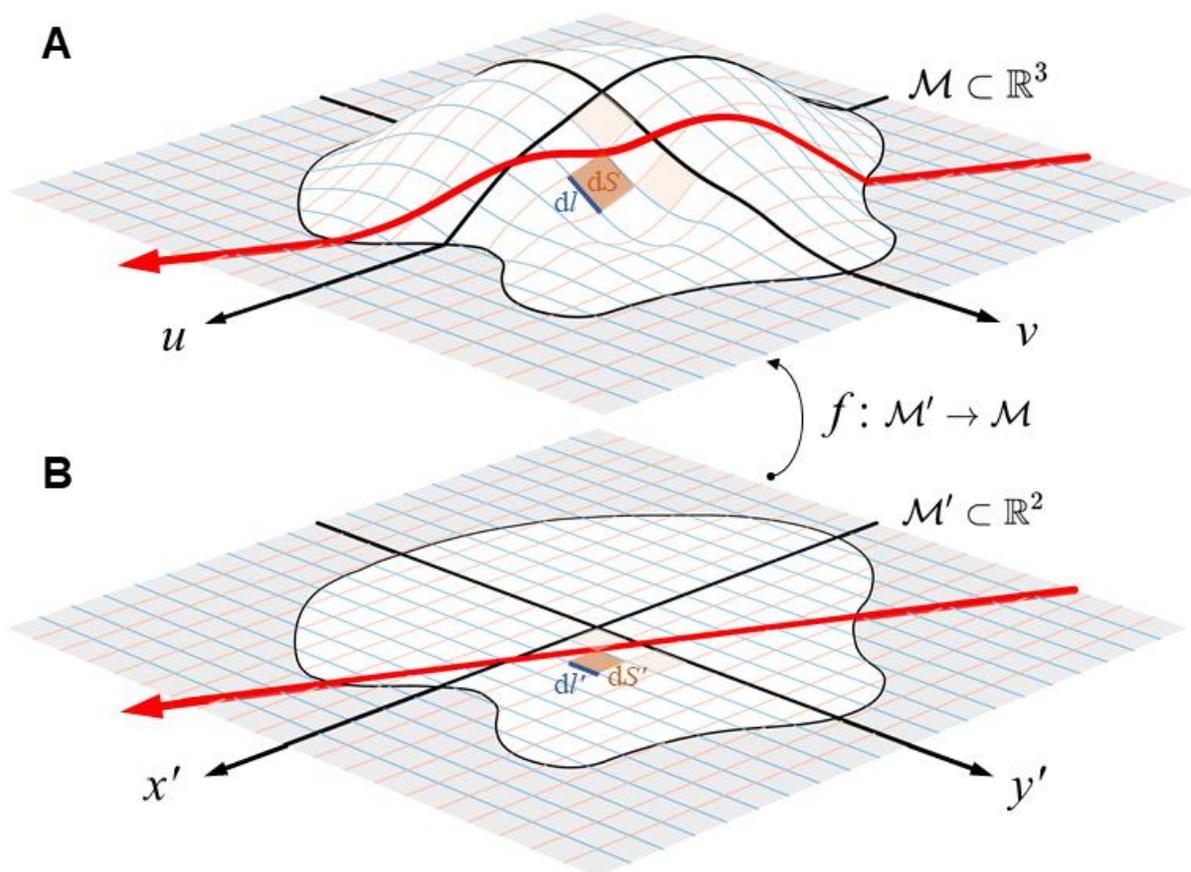

**Fig. 1. The conformal mapping between manifolds.** (A) A light beam crossing a curved two-dimensional manifold $\mathcal{M}$ embedded in $\mathbb{R}^3$. (B) A light beam crossing a flat two-dimensional manifold $\mathcal{M}'$ in $\mathbb{R}^2$. The manifold $\mathcal{M}$ is $uv$-parameterized and both manifolds are plotted with coordinate grid. One can obtain the manifold $\mathcal{M}$ in (A) from $\mathcal{M}'$ in (B) through a certain analytic or numerical mapping $f: \mathcal{M}' \to \mathcal{M}$.

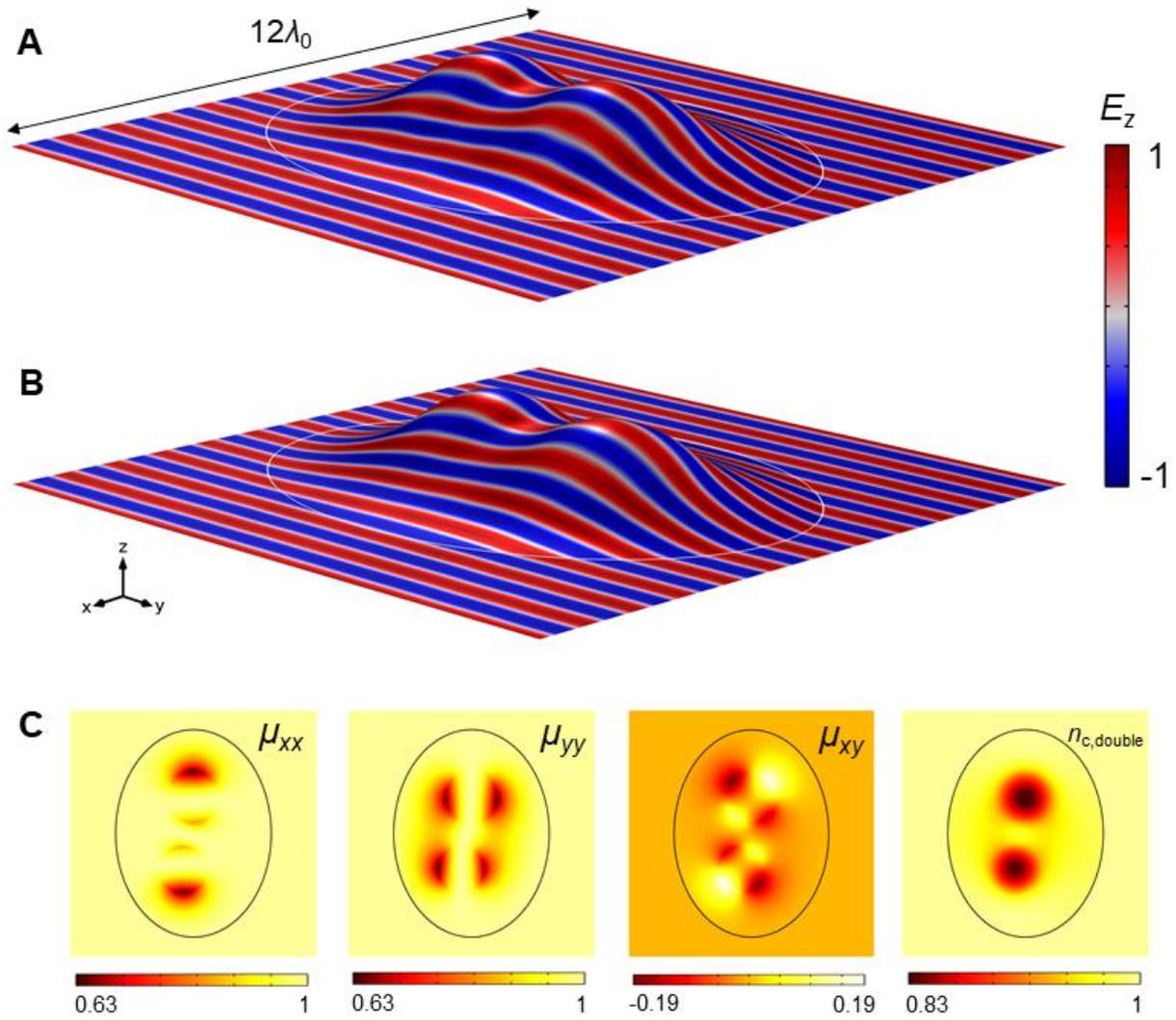

**Fig. 2. The field and medium distribution for cloaks.** Normalized electric field distribution of surface electromagnetic wave cloaks achieved by (A) anisotropic relative permeability and (B) isotropic refractive index. (C) Components of anisotropic relative permeability, $\mu_{xx}$, $\mu_{yy}$ and $\mu_{xy}$, applied in (A) and isotropic refractive index $n_{c,\text{double}}$ applied in (B). The excitation is a $z$-polarized plane wave with a magnitude of $|E_z| = 1$ V/m; and the wavelength in free-space is $\lambda_0 = 20$ mm. The bump with a height of $1.25\lambda_0$ is located in the center of the square waveguide with a width of $12\lambda_0$. For the elliptical boundary, the semi-minor and semi-major axis length are $a = 3.75\lambda_0$ and $b = 5\lambda_0$, respectively, along with $x$- and $y$-axes.

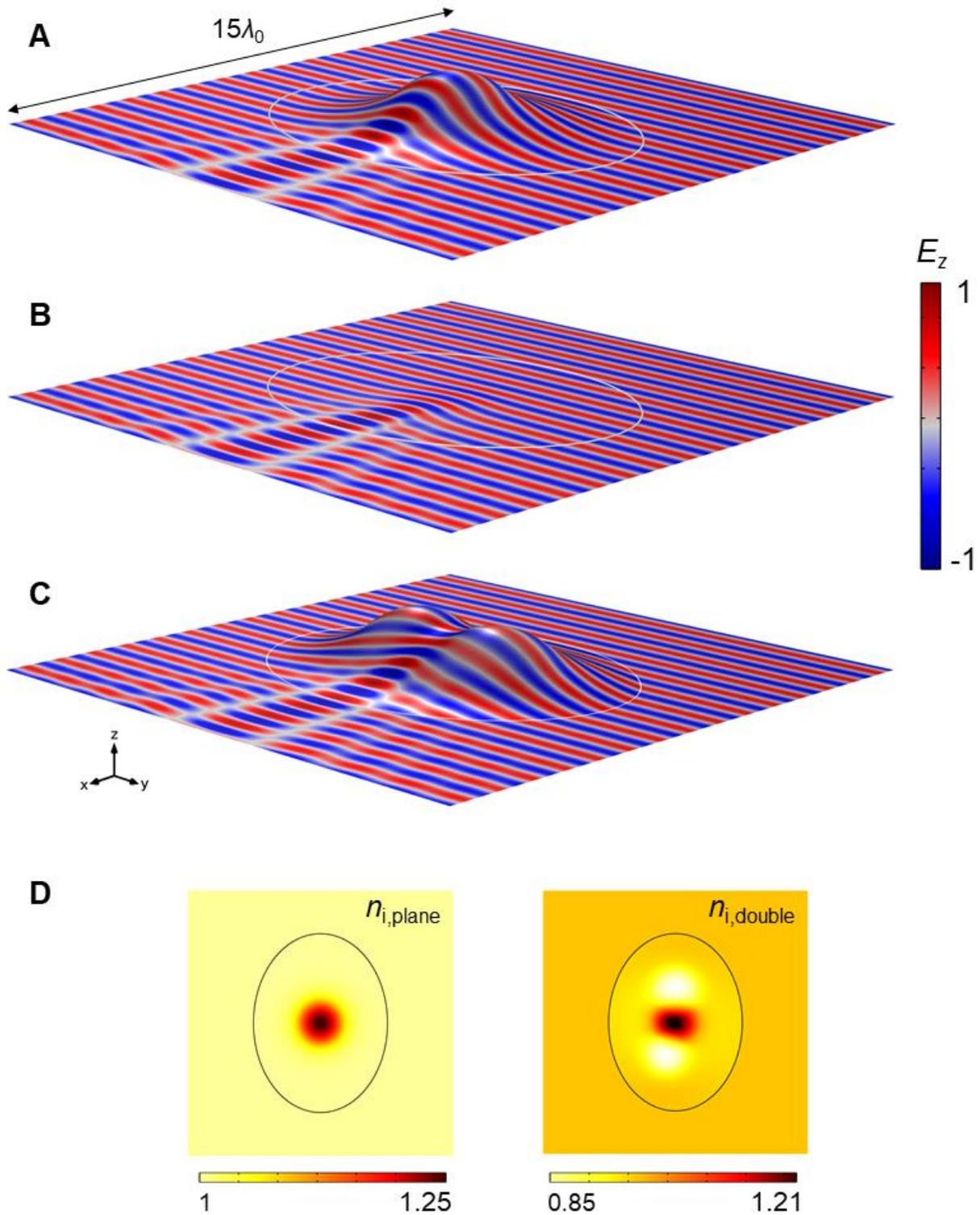

**Fig. 3. The field and medium distribution for illusions.** Normalized electric field distribution of surface electromagnetic wave scattering. (A) Scattering on the single-camelback bump when filled with homogeneous medium. (B) Illusion of the single-camelback bump appearing on the plane. (C) Illusion of the single-camelback bump appearing on the double-camelback bump. (D) Isotropic refractive indices: $n_{i,\text{plane}}$ for the elliptic region in (B) and $n_{i,\text{double}}$ for the double-camelback bump in (C). The elliptical base profiles in (A), (B) and (C) are the same.

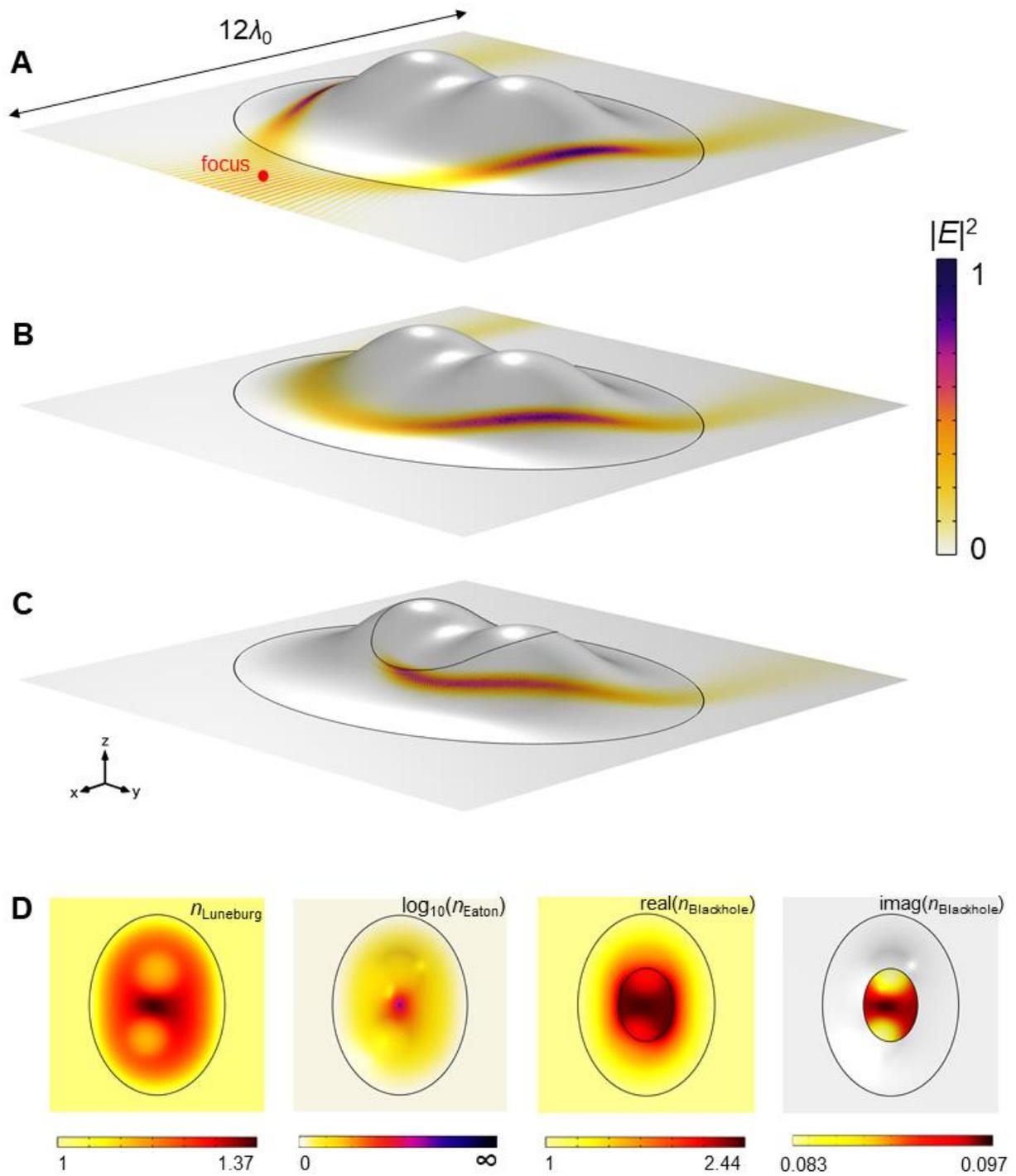

**Fig. 4. The field and medium distribution for devices.** Normalized electric field distribution on surface electromagnetic wave devices. (A) Luneburg lens; (B) Eaton lens; and (C) Black hole. Gaussian beam is applied to demonstrate their functions. (D) Isotropic refractive indices; $n_{\text{Luneburg}}$ for Luneburg lens in (A), decimal logarithm of $n_{\text{Eaton}}$ for Eaton lens in (B), real and imaginary part of $n_{\text{Blackhole}}$ for black hole in (C).

# Supplementary Materials for

- **Controlling Electromagnetic Surface Waves with Conformal Transformation Optics**

Xiaoyu Zhao *et al.*

*Corresponding author. Email: tydong@mail.xjtu.edu.cn.

**This PDF file includes:**

Figs. S1 to S5

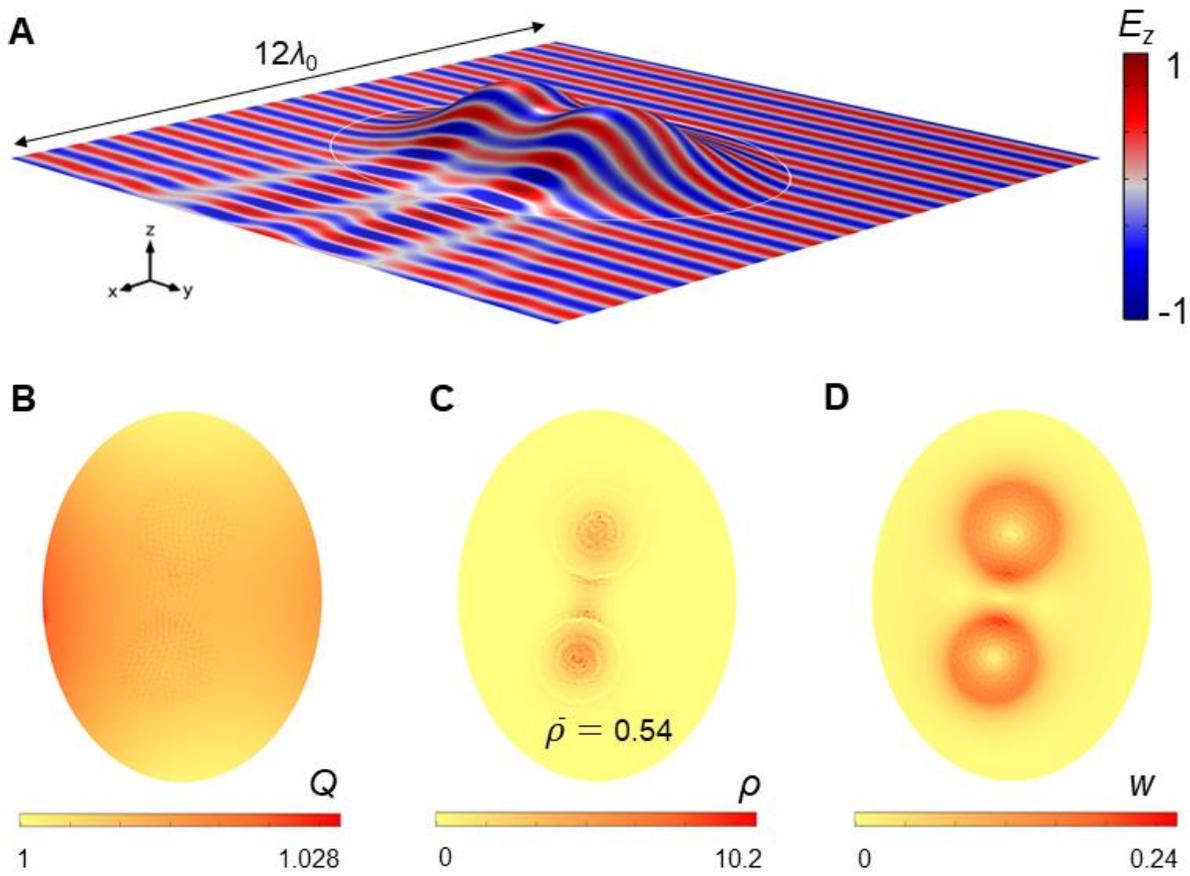

**Fig. S1.** (A) Normalized electric field distribution of surface electromagnetic wave scattering on double-camelback bump when filled with homogeneous medium. (B) Quasi-conformal ratio $Q$ of the mapping applied to design the cloak shown in Fig. 2A. (C) Curvature index $\rho$. (D) Wavelength index $w$.

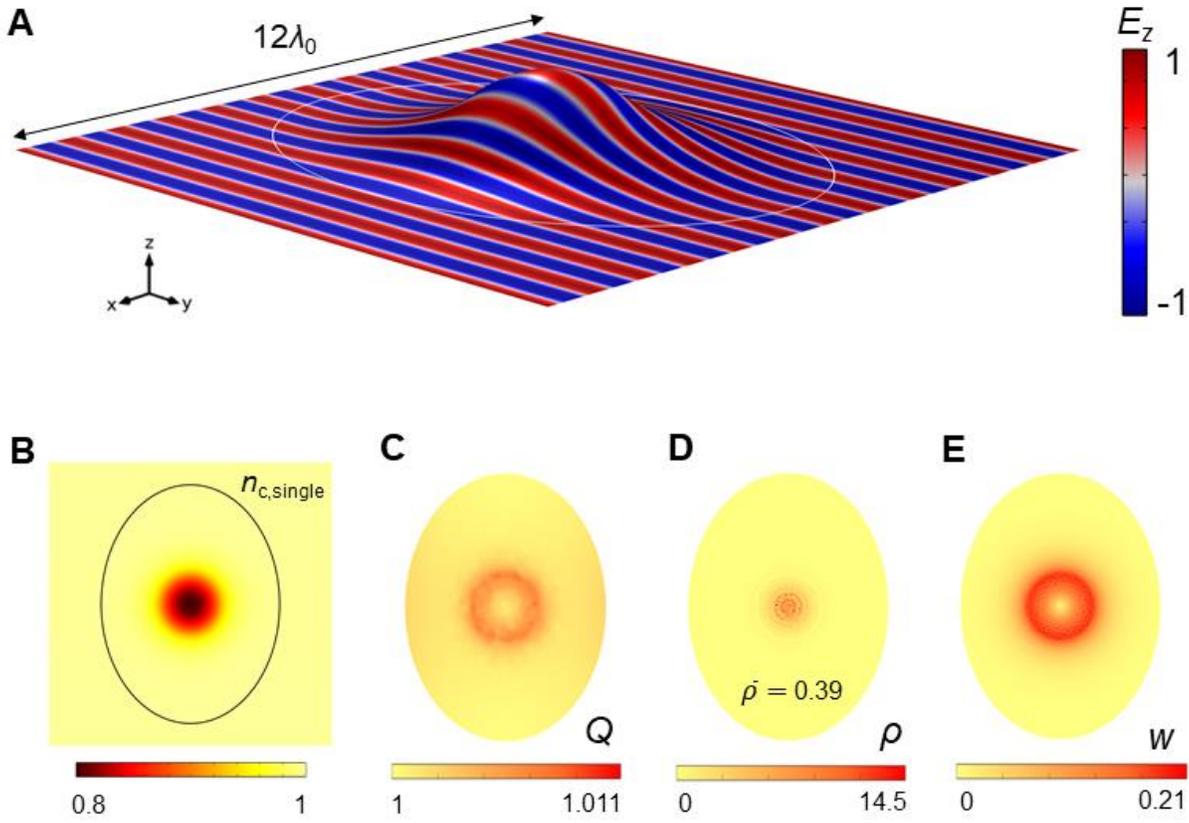

**Fig. S2.** (A) Normalized electric field distribution of the surface electromagnetic wave cloak on single-camelback bump achieved by (B) isotropic refractive index $n_{c,\text{single}}$. (C) Quasi-conformal ratio $Q$ of the mapping applied to design the cloak shown in (A). (D) Curvature index $\rho$. (E) Wavelength index $w$.

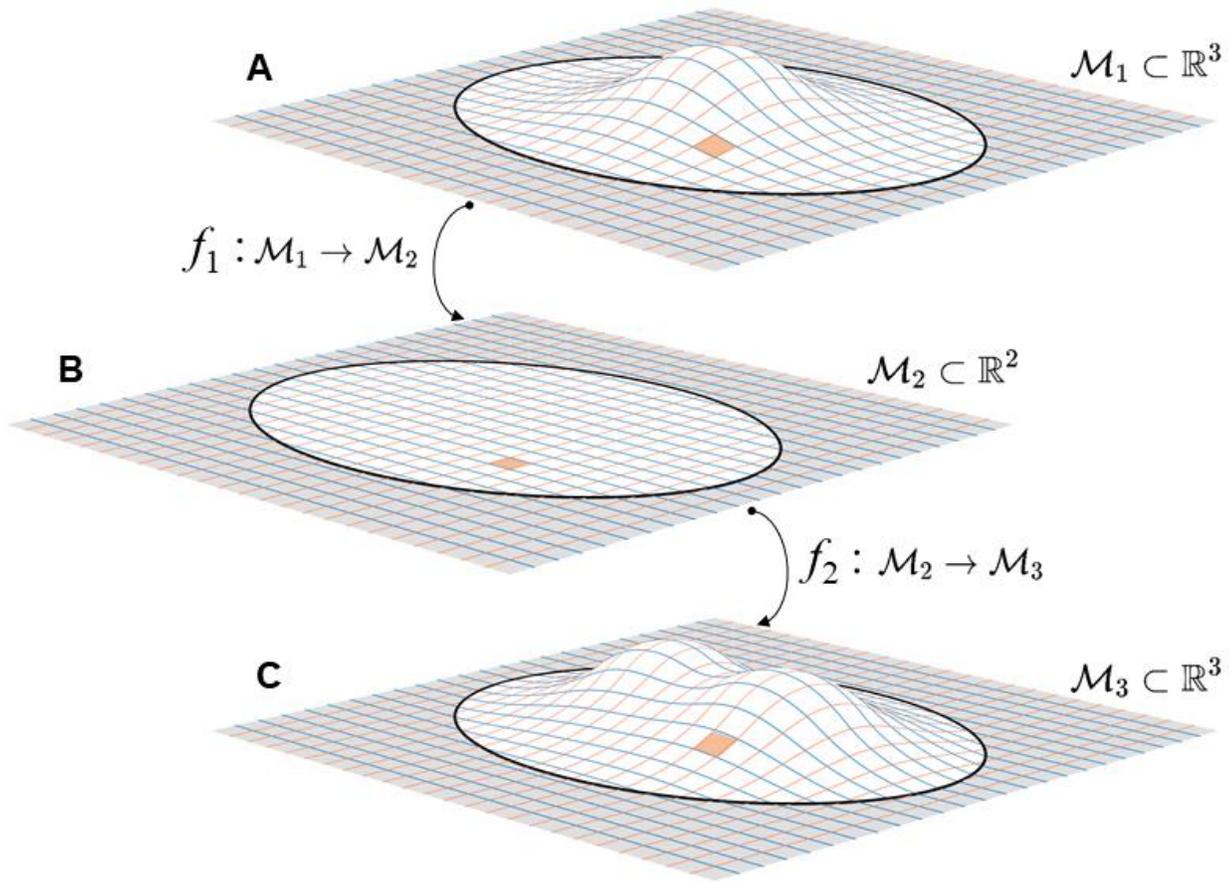

**Fig. S3.** A quasi-conformal mapping between two manifolds embedded in $\mathbb{R}^3$ constructed by cascading two mappings between $\mathbb{R}^3$ and $\mathbb{R}^2$. (A) A single-camelback manifold $\mathcal{M}_1$ embedded in $\mathbb{R}^3$. (B) The plane region $\mathcal{M}_2$ in $\mathbb{R}^2$ mapped from $\mathcal{M}_1$ through mapping $f_1$. (C) The double-camelback manifold $\mathcal{M}_3$ embedded in $\mathbb{R}^3$ mapped from $\mathcal{M}_2$ through mapping $f_2$.

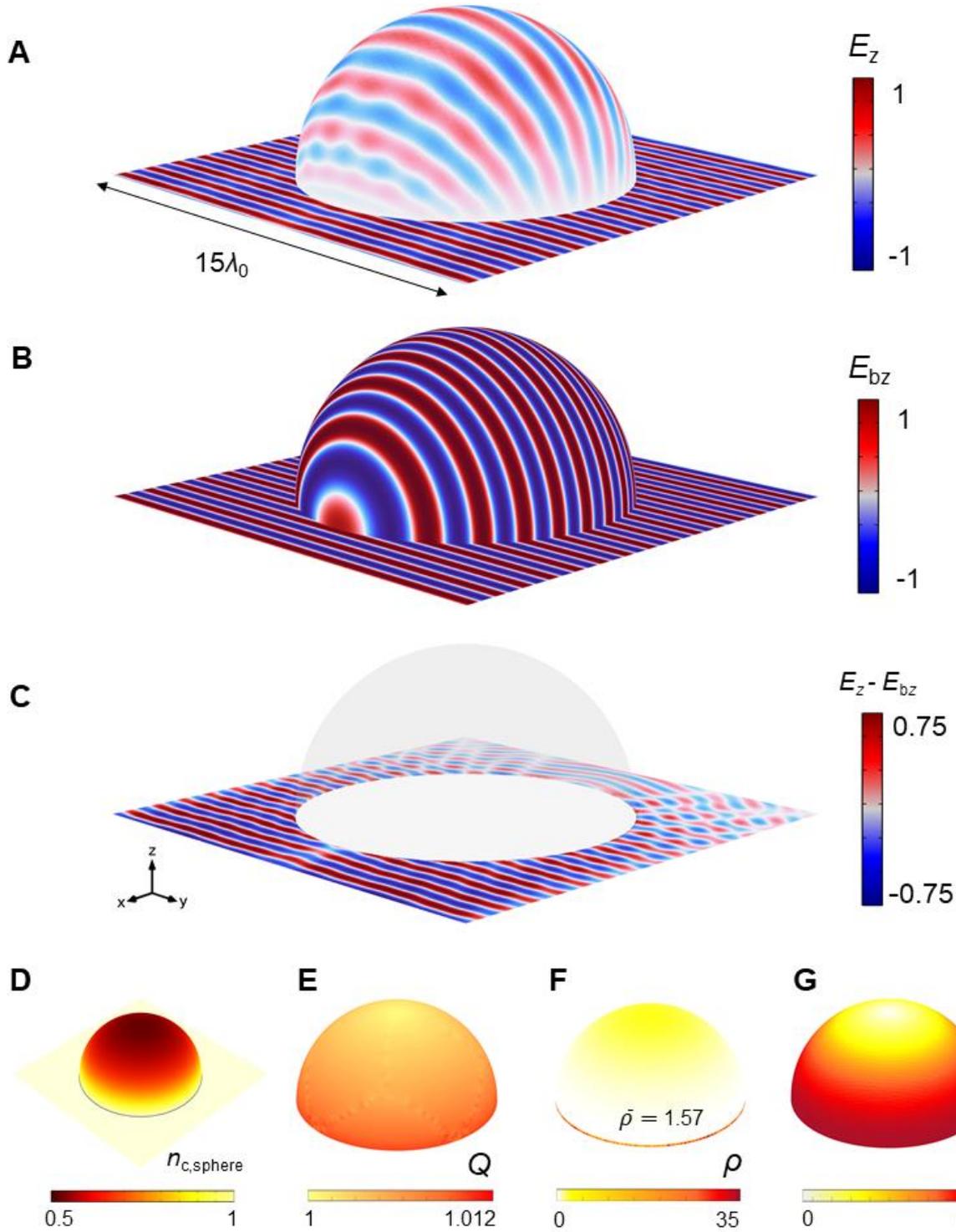

**Fig. S4.** (A) Normalized electric field $E_z$, (B) background field $E_{bz}$ and (C) scattering field $E_z - E_{bz}$ of the hemisphere surface wave cloak achieved by (D) isotropic refractive index $n_{c,\text{sphere}}$. (E) Quasi-conformal ratio $Q$ of the mapping applied to design the cloak shown in (A). (F) Curvature index $\rho$. (G) Wavelength index $w$. The radius of the hemisphere is $5\lambda_0$.

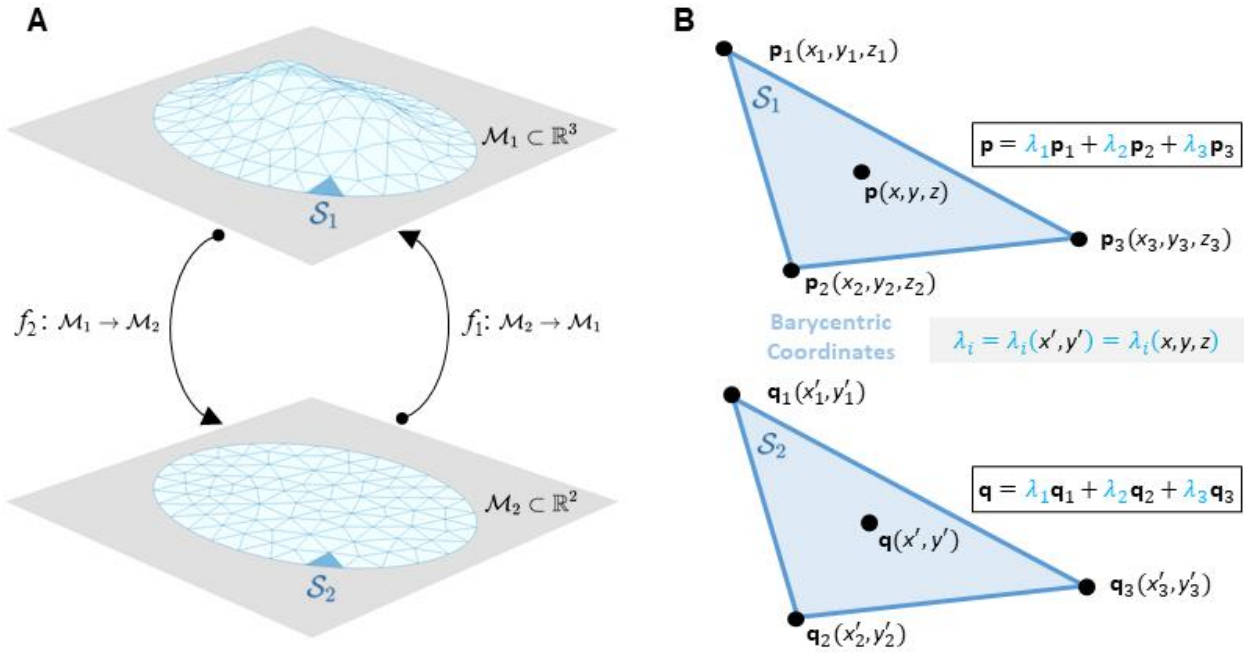

**Fig. S5.** (A) Simplices $\mathcal{S}_1$ and $\mathcal{S}_2$ as triangle elements in the mesh of double-camelback manifold $\mathcal{M}_1$ embedded in $\mathbb{R}^3$ and the region $\mathcal{M}_2$ in $\mathbb{R}^2$, related by quasi-conformal mappings $f_1$ and $f_2$. (B) The same barycentric coordinates on simplices $\mathcal{S}_1$ and $\mathcal{S}_2$.